\documentclass{aastex631}
\usepackage{amsmath}
\usepackage{booktabs}
\usepackage{multirow}
\usepackage{array}
\usepackage{makecell}
\usepackage[utf8]{inputenc}
\usepackage{dsfont}
\begin{document}

\title{Architecture of  Tianyu Software: Relative Photometry as a Case Study}

\author[0000-0001-5295-1682]{Yicheng Rui}
\altaffiliation{ruiyicheng@sjtu.edu.cn}
\affiliation{State Key Laboratory of Dark Matter Physics, Tsung-Dao Lee Institute \& School of Physics and Astronomy, Shanghai Jiao Tong University, Shanghai 201210, China}
\author[0009-0000-9341-3442]{Yifan Xuan}
\affiliation{State Key Laboratory of Dark Matter Physics, Tsung-Dao Lee Institute \& School of Physics and Astronomy, Shanghai Jiao Tong University, Shanghai 201210, China}
\author{Shuyue Zheng}
\affiliation{State Key Laboratory of Dark Matter Physics, Tsung-Dao Lee Institute \& School of Physics and Astronomy, Shanghai Jiao Tong University, Shanghai 201210, China}
\author{Kexin Li}
\affiliation{State Key Laboratory of Dark Matter Physics, Tsung-Dao Lee Institute \& School of Physics and Astronomy, Shanghai Jiao Tong University, Shanghai 201210, China}
\author[0000-0003-1535-5587]{Kaiming Cui}
\affiliation{Department of Physics, University of Warwick, Gibbet HIll Road, Coventry CV4 7AL, UK}
\affiliation{Centre for Exoplanets and Habitability, University of Warwick, Gibbet Hill Road, Coventry CV4 7AL, UK}
\author[0000-0001-8424-1079]{Kai Xiao}
\affiliation{School of Astronomy and Space Science, University of Chinese Academy of Sciences, Beijing 100049, People's Republic of China}
\affiliation{Institute for Frontiers in Astronomy and Astrophysics, Beijing Normal University, Beijing, 102206, China}
\author[0000-0001-6637-6973]{Jie Zheng}
\affiliation{CAS Key Laboratory of Optical Astronomy, National Astronomical Observatories, Chinese Academy of Sciences,
Beijing, China}
\author{Jun Kai Ng}
\affiliation{State Key Laboratory of Dark Matter Physics, Tsung-Dao Lee Institute \& School of Physics and Astronomy, Shanghai Jiao Tong University, Shanghai 201210, China}
\author[0000-0003-0292-2773]{Hongxuan Jiang}
\affiliation{State Key Laboratory of Dark Matter Physics, Tsung-Dao Lee Institute \& School of Physics and Astronomy, Shanghai Jiao Tong University, Shanghai 201210, China}
\author{Fabo Feng}
\altaffiliation{ffeng@sjtu.edu.cn}
\affiliation{State Key Laboratory of Dark Matter Physics, Tsung-Dao Lee Institute \& School of Physics and Astronomy, Shanghai Jiao Tong University, Shanghai 201210, China}
\author{Qinghui Sun}
\affiliation{State Key Laboratory of Dark Matter Physics, Tsung-Dao Lee Institute \& School of Physics and Astronomy, Shanghai Jiao Tong University, Shanghai 201210, China}



\begin{abstract}
 Tianyu telescope, an one-meter robotic optical survey instrument to be constructed in Lenghu, Qinghai, China, is designed for detecting transiting exoplanets, variable stars and transients. It requires a highly automated, optimally distributed, easily extendable, and highly flexible software to enable the data processing for the raw data at rates exceeding 500MB/s. In this work, we introduce the architecture of the Tianyu pipeline and use relative
 photometry as a case to demonstrate its high scalability and efficiency. This pipeline is tested on the data collected from Muguang observatory and Xinglong observatory. The pipeline demonstrates high scalability, with most processing stages increasing in throughput as the number of consumers grows. Compared to a single consumer, the median throughput of image calibration, alignment, and flux extraction increases by 41\%, 257\%, and 107\% respectively when using 5 consumers, while image stacking exhibits limited scalability due to I/O constraints.  In our tests, the pipeline was able to detect two transiting sources. Besides, the pipeline captures variability in the light curves of nine known and two previously unknown variable sources in the testing data. Meanwhile, the differential photometric precision of the light curves is near the theoretical limitation. These results indicate that this pipeline is suitable for detecting transiting exoplanets and variable stars. This work builds the fundation for further development of Tianyu software.
 Code of this work is available at \url{https://github.com/ruiyicheng/Tianyu_pipeline}.
\end{abstract}

\keywords{Astronomy software(1855) --- Photometry(1234)}


\section{INTRODUCTION}
\label{sec:intro}  % \label{} allows reference to this section

Tianyu is an one-meter robotic optical survey telescope set to be constructed in Lenghu, Qinghai, China \citep{Feng24}. With its multi-cadence observation modes spanning from sub-second to monthly intervals, Tianyu is designed to detect a wide range of astronomical phenomena, including long-period exoplanets, solar system objects, variable stars, and transient events such as gamma-ray bursts (GRBs), supernovae, Intermediate Luminosity Optical Transients (ILOTs) and kilonovae.

The success of the Tianyu project hinges on the design of a robust software system. A well-architected system is crucial for automating the generation of high-level scientific products and ensuring efficient process management. For instance, the achievements of the Zwicky Transient Facility (ZTF) \citep{Masci19} are largely due to its advanced software, which automates the daily production of supernova alerts, enabling rapid follow-up observations. This capability is also essential for Tianyu, particularly in detecting long-period transiting exoplanets.

A key component of such automation is an architecture that ensures efficient and reliable process management. ZTF employs a Kafka-based system \citep{Kreps2011KafkaA} for handling data streams. Besides, the Large Synoptic Survey Telescope (LSST) utilizes a shared-nothing distributed database for data management \citep{wang11}. The middleware of LSST, which is used for process management, is based on this database. The application layer of LSST software is built based on this middleware, which contains the data processing pipeline \citep{juri17}. This sophisticated architecture enables LSST to handle the immense data throughput of over 15 TB per night generated by its camera. % This paper presents the design and implementation of the Tianyu software system, drawing insights from these proven frameworks while addressing the specific needs of the Tianyu project.
To achieve full automation of observation and data processing like ZTF and LSST, Tianyu requires a architecture for process management which is similar to these projects. The data processing pipeline and the telescope control software of Tianyu  are based on the architecture.

%Besides, a key component of the Tianyu software is relative photometry for intra-night images, which is essential for detecting exoplanets. This technique is not utilized by the ZTF due to its wide-field survey strategy \citep{Masci19}, nor is it a focus of the Kepler mission, which benefits from the stability of space-based observations \citep{Morris20}. Achieving absolute photometric calibration with a precision exceeding 1\%—the typical signal strength of a transiting Jupiter—is particularly challenging for ground-based telescopes due to atmospheric variability \citep{Padmanabhan08}. As a result, relative flux extraction is a more practical and effective approach for identifying transiting exoplanets. Furthermore, high-precision relative flux measurements are also valuable for detecting variable stars.
  The Tianyu data processing pipeline incorporates relative photometry for intra-night images, enabling precise flux extraction despite atmospheric variability. This method is essential for detecting exoplanets and is also valuable for studying short-timescale variable stars, such as EW-type eclipsing binaries and stellar flares.

% Meanwhile, forced photometry is required to detect signals in dark sources. ZTF provided a user-friendly tool to enable users to get the forced photometry light curves for up to 1500 positions in the sky using a web request \citep{Masci23}. However, getting the forced photometry results of all the sources of the deep template images would be helpful

This paper provides a comprehensive introduction to the architecture of the Tianyu software system and presents a preliminary relative photometry pipeline developed based on this architecture. Real observational data from multiple telescopes are employed to evaluate the pipeline's performance. %Section \ref{sec:req} outlines the software requirements for the Tianyu project. 

Section \ref{sec:software} outlines big data challenge for Tianyu and provides a detailed description of the different components in Tianyu software, including the architecture, database system and file system. Section \ref{sec:pipelines} gives a detailed description of the relative photometry pipeline based on this architecture. In Section \ref{sec:test}, we assess the performance of the software system and the relative photometry pipeline using real-world data. Finally, Section \ref{sec:condis} summarizes the contributions of this study and outlines future directions for the Tianyu software system.

 % There will be a series of paper on the Tianyu software. In Paper I (this paper), we present the architecture and light curve component of Tianyu software; In paper II, we present a event-based simulator for ground-based time-domain surveys; In paper III, we will describe the the data challenge based on this simulator; In paper IV, we will show the results of data challenge; In paper V, we will present the scheduling and controlling system of Tianyu.

\section{Software system}
\label{sec:software}
\subsection{Big data challenge for Tianyu}
\label{sec:req}
Tianyu's data generation capability, with raw data at rates exceeding 500MB/s from multiple instruments, would lead to difficulties when using traditional data processing pipeline.  These pipelines often require manual intervention to manage dependencies between processes, making them inefficient for large-scale survey projects that cover multiple sky regions each night.  Furthermore, traditional pipelines are typically designed for centralized computing environments, making them unsuitable for remote locations with limited infrastructure. Moreover, traditional architectures are rigid and difficult to extend, making it challenging to incorporate new functionalities or update algorithms without significant restructuring. Resource management is also inflexible, with traditional systems unable to dynamically allocate computing resources across heterogeneous environments, such as Windows-based telescope control and Linux-based data processing. Additionally, traditional pipelines rely on simplistic file-based storage systems, which are inadequate for managing the diverse range of scientific data Tianyu will generate. 
Therefore, Tianyu requires the development of a high-performance software system with following features. 

\begin{itemize}
    \item \textbf{Highly automated:}  In general, dependence between processes of software are managed manually. For example, one would be able to execute data process pipeline after the observation is finished. One still need to input the metadata of observation to make the data reduction pipeline run smoothly. For Tianyu, automation of the processes are required because the dependence of a survey project is too complicated for human to manage. About 50 sky regions would be covered each night. The system should automatically schedule observations, acquire, and process data. Meanwhile, the system should be able to release alerts automatically in emergent situation like power failures. 
    
    \item \textbf{Optimally distributed:} The Tianyu telescope's remote location on Saishiteng Mountain creates challenges for data processing and storage. 
    Because deploying a large server is difficult due to the terrain, alternative solutions are needed. To overcome the limited network bandwidth, data compression and image stacking in-situ are necessary to reduce the amount of data that needs to be transmitted.

Individual nodes in a computing system typically have limited storage and processing capabilities, necessitating the adoption of distributed systems for both storage and database management. These systems enable efficient data storage and retrieval across multiple nodes, while ensuring scalability in computational resources to accommodate growing demands. In software systems, parallelism plays a critical role in enhancing performance and scalability. There are two primary types of parallelism: (1) pipeline parallelism, where different stages of a process are executed simultaneously, allowing scalability up to the pipeline depth; and (2) data parallelism, where different subsets of input data are processed concurrently, enabling scalability proportional to the volume of data. For the Tianyu software, implementing both types of parallelism is essential to achieve its scientific objectives. This dual approach allows the system to efficiently process large datasets and meet the stringent time requirements of producing results for multiple scientific targets within a single day.

    \item \textbf{Easily extendable:} Due to the limitations of the foundation and human resources, Tianyu will only realize its fundamental functionality in its current stage. The software designed for Tianyu should be extendable, meaning it must be easy to add new functions and revise algorithms.

    \item \textbf{Highly flexible:} Tianyu will utilize multiple telescopes and operating systems, such as Windows for telescope control and Linux for data processing pipelines. Different components of the 
software require varying resources, with image stacking requiring large RAM and classification algorithms necessitating GPU acceleration. To accommodate these diverse needs, the software must be able to dynamically allocate resources to specific computers and accomplish different types of missions.

Meanwhile, Tianyu will generate different kinds of scientific data ranging from light curve to truncated image. A relational database system is needed for users to obtain different kinds of data easily.
\end{itemize}

\subsection{Architecture of Tianyu software system}
\label{sec:architecture}
Tianyu software uses a publisher-consumer based model for process management. A demonstration of architecture is shown in Fig. \ref{fig:architecture}. Tianyu software consists of three parts: process (defined in Appendix \ref{sec:concepts}) publisher, process manager and process consumer. 

When a process  is published, the publisher would insert command of the process, parameters of this process, site of the consumer (defined in Appendix \ref{sec:concepts}), group of the consumer (defined in Appendix \ref{sec:concepts}) and the dependence of the process into the database. The information provided should be sufficient for consumer to execute as long as the dependent processes are been successfully executed. Different groups are needed because different processes need different resources, e.g. image stacking needs large RAM; object classification needs GPU; observation needs telescope and camera. This would enable us to treat some special operation like capture image as regular processes in specific groups. Meanwhile, this would enable fine allocation of resources combined with container orchestration techniques.

Process manager would scan the database of process list periodically. Once all of the dependence of a process is successfully executed, process manager would push the command and parameter of the process into queue broker. Every consumer group have its unique queue. Finally, process manager would refresh the status of this process once they are successfully pushed into the command queue. \textsc{RabbitMQ}\footnote{\url{https://www.rabbitmq.com/}} is the queue broker used in this work.

Process consumer receive message from queue broker. Once it receive a command, it execute the corresponding back end software, e.g. image stacking.  When the consumer succeeded or failed to execute the command, the consumer would register the status of the process.

This architecture enable every consumer using the same code. The only difference between consumers is the sites and groups of them. This would make the pipeline easy to develop, update and deploy. Meanwhile, this architecture would easily achieve both kinds of parallelism as mentioned in Section \ref{sec:req} by deploying more of them.

\begin{figure}
    \centering
    \includegraphics[width=1\linewidth]{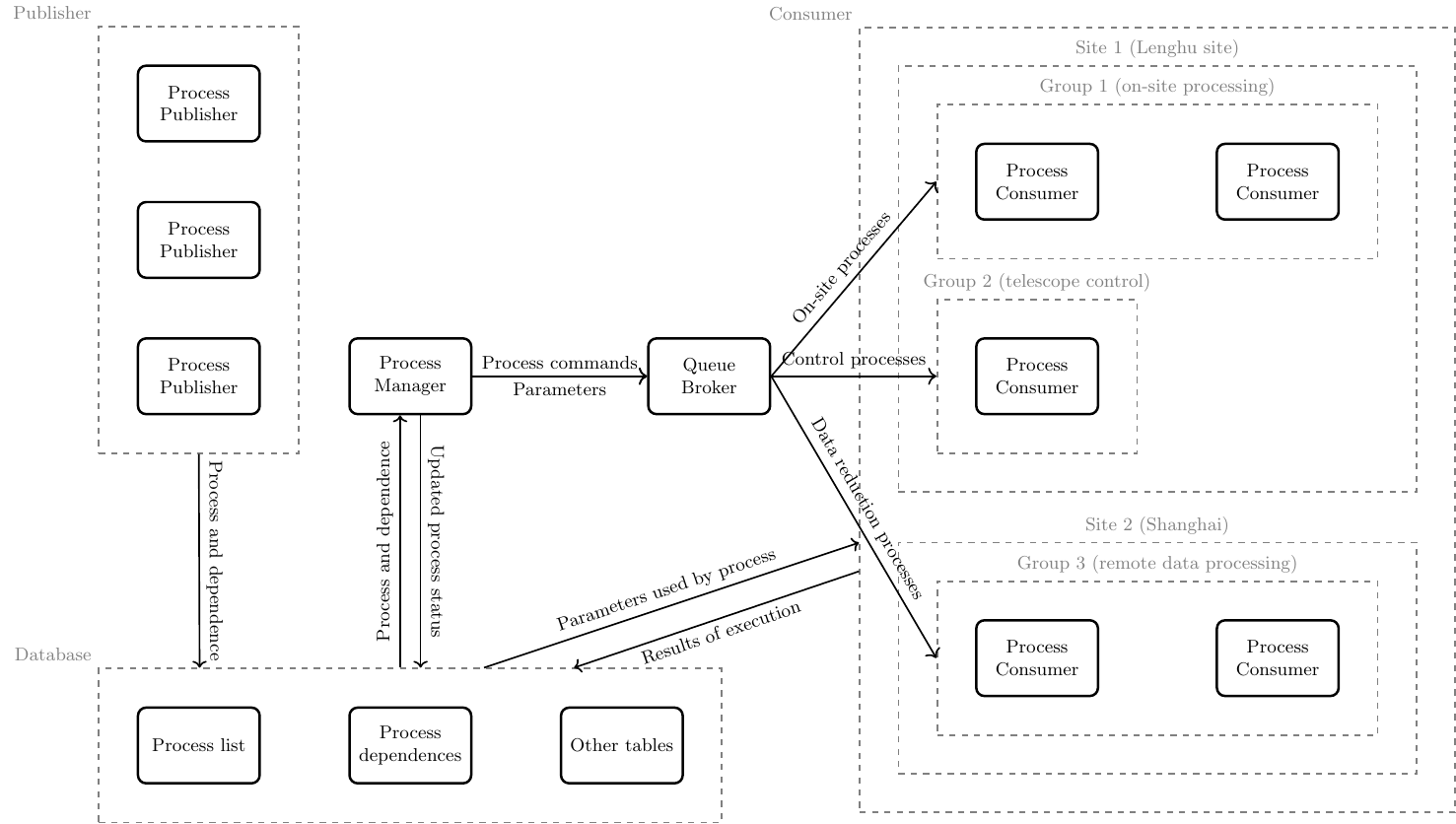}
    \caption{Architecture of Tianyu software. See details in Section \ref{sec:architecture}.}
    \label{fig:architecture}
\end{figure}
\subsection{Database system of Tianyu software}
\label{sec:database}

In Tianyu software, a MySQL-based database \footnote{\url{https://www.mysql.com/}} is used to store and manage the telescope's observational data, intermediate data and result data. Because the columns of the database is still updating, a detailed description of the database will be described in the github repository. Schemas of the tables of the database used in this work is shown in Appendix \ref{sec:dbs}.

\begin{figure}
    \centering
    \includegraphics[width = 0.8\linewidth]{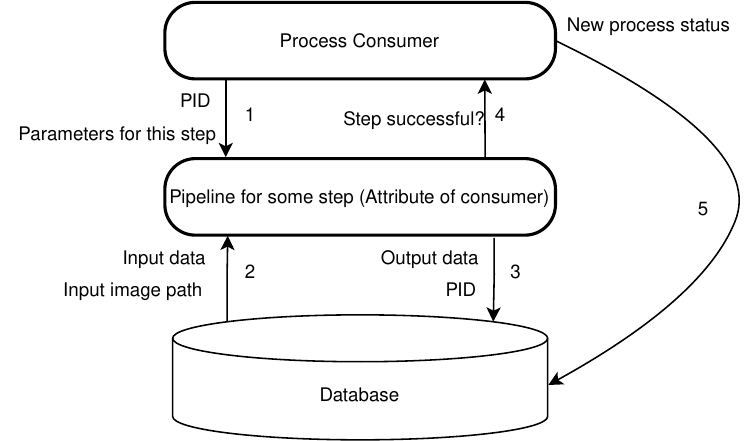}
    \caption{A general coupling relationship between the pipeline and the database. See details in Section \ref{sec:database}.}
    \label{fig:couplingdbpipe}
\end{figure}

 Database not only provides a rapid and flexible way for users to query the data products, but is coupled with data process pipeline as well. There are two kinds of tables in the database. The first kind of table is fixed during the data processing . They provide data that would be used by the pipeline. For example, \textit{obs\_site} would provide the information of the observational site, which is fixed when processing data. Meanwhile, records of the second kind of table is generated or changed during the data reduction. They record the results and \textbf{process\_id} of a process, making it easier for the next step of the pipeline to locate the input data. For example, after resolving the sources (defined in Appendix \ref{sec:concepts}) in template, pipeline would record the information of resolved sources in \textit{tianyu\_source\_position} and record the map between template and sky in \textit{process\_id}. When selecting reference stars, the pipeline can use that \textit{process\_id} as a handle to obtain the information of resolved sources. A table can have more than one \textit{process\_id}. It would enable recording the handle for multiple processes. For example, table \textit{img} would record \textit{birth\_process\_id} and \textit{align\_process\_id} to enable processes to obtain the image generation results, i.e. captured, stacked or calibrated image, and alignment results respectively. A general coupling relationship between the pipeline and the database is shown in Fig \ref{fig:couplingdbpipe}. After a consumer receive the command of process from the queue broker, the consumer would interact with the database as follows:

\begin{enumerate} 
\item The consumer parses the information encoded in JSON and inputs it into the corresponding pipeline program, along with the process's \textit{process\_id} (PID). \item The pipeline program retrieves the necessary data from the database based on the provided parameters. These parameters may include PIDs of data generated by other processes or pipeline configuration details. It then executes the pipeline using the retrieved data and specified parameters. \item The output data and the PID of this process are recorded in the database. 
\item The pipeline reports to the process consumer whether steps 2 and 3 were successfully executed. 
\item The consumer updates the process status based on the success or failure of the pipeline execution. 
\end{enumerate}

The Tianyu database employs transactions to ensure data consistency. Specifically, the publication of \textit{process\_id} and its dependency relationships are encapsulated within a single transaction. This approach prevents transient states where a process is published, but its dependencies are not yet recorded in the database. Such states could lead the process manager to incorrectly enqueue the process into the queue broker due to the absence of its dependencies. By combining these operations into a single transaction, the process publisher ensures the integrity and completeness of the dependency information.

\subsection{File system of Tianyu}

Recording all utilized information in the database may seem appealing. However, reading and writing image data to and from the database is highly time-consuming. Consequently, a file system remains necessary for storing image files. Additionally, software such as \textsc{Source Extractor} \citep{bertin96} and textsc{SCAMP} \citep{bertin06} cannot directly access data from the database. Therefore, a file system is also required to manage configuration files and input catalogs for these tools.

%The detailed architecture of file system is shown in Fig. \ref{fig:filesys} of Appendix \ref{sec:detailedfs}.
For each site, there is a fixed root directory to restore the files. This root directory is recorded in table \textit{data\_process\_site} of the database. An image have fixed path relative to the site root directory. \textcolor{black}{The paths of the image files are encoded in their metadata, which is stored in the database. When a user or pipeline requires a batch of images, e.g.  all images taken in a single night for a given target, they must first query the relevant entries from the database. This process is straightforward, as the image table includes foreign keys linking it to observation tables. Subsequently, a function is used to convert the retrieved database entries into the corresponding file paths.} This design would enable image transfer between different remote sites. 

\section{Relative photometry pipelines}
\label{sec:pipelines}

\begin{figure}
    \centering
    \includegraphics[width = \linewidth]{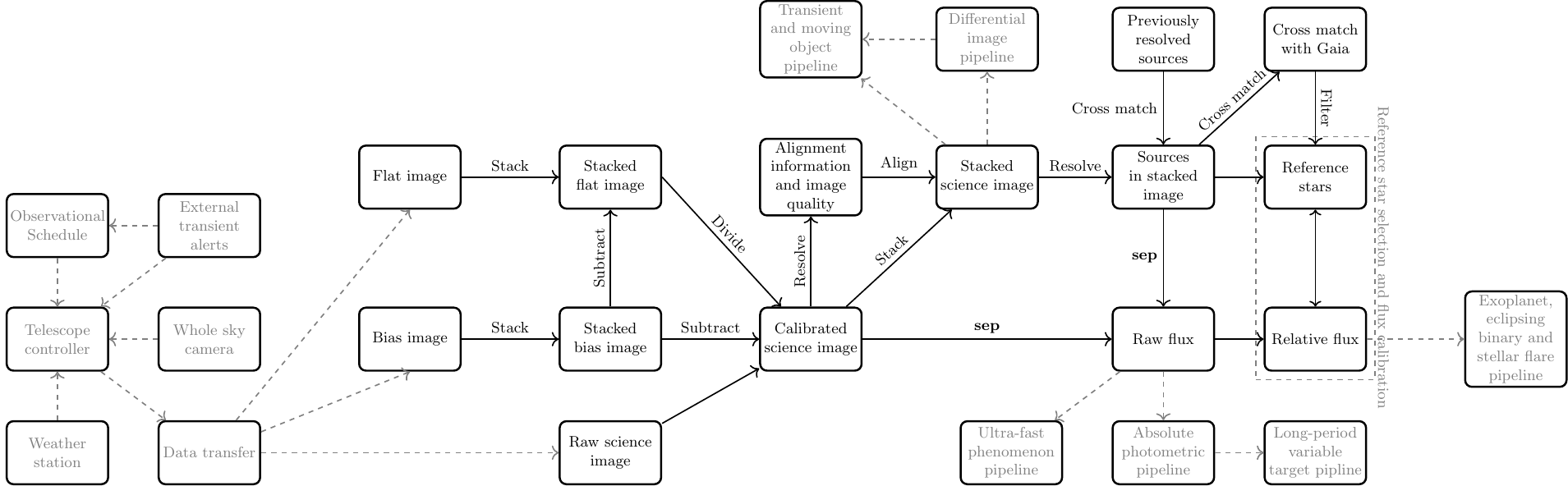}
    \caption{Data flow of the Tianyu pipeline. Blocks with black labels represent stored data entities generated by this pipeline, as illustrated in Fig. \ref{fig:couplingdbpipe}. Arrows indicate the direction of data flow and associated processing operations. Blocks with gray labels and dashed arrows denote components planned for implementation in future work. Further details are provided in Section \ref{sec:pipelines}.}
    \label{fig:dataflow}
\end{figure}

To validate the architecture proposed in this study, a preliminary pipeline for extracting relative flux has been developed. This pipeline is also a crucial component for detecting transiting exoplanets and variables. Although functional, there is significant potential for further optimization. Additional steps, such as absolute flux calibration and differential image, are currently under development and will be presented in future work. A process publisher would register the process and dependence of all the steps given the pre-registered observations' (defined in Appendix \ref{sec:concepts}) process id and sky's (defined in Appendix \ref{sec:concepts}) process id. The data flow of Tianyu pipeline is illustrated in Fig. \ref{fig:dataflow}. The dependency relationships between different processes published by the publisher in the pipeline follow the reverse direction of the arrows; that is, the initiation of subsequent data product generation is triggered by the completion of the process that produces the original data. The triggering mechanisms between different components of the pipeline are described in Section \ref{sec:architecture}. Image products are stored in the file system, while other data, such as light curves, are stored in the database. The details of each step in the current relative photometry pipeline are presented in the following sections.

\subsection{Image calibration} Image calibration is essential to correct for the uneven response to an evenly distributed light source and to compensate for bias currents. In this calibration pipeline, the stacked super-bias frame \( B \) is obtained by averaging multiple bias images. The process for generating the master flat frame is more complex. First, the stacked super-bias frame is subtracted from each flat field image \( F_i \), where \( i \) denotes the index of the flat image. The resulting debiased flat image is given by \( \tilde{F_i} = F_i - B \). Next, each debiased flat image \( \tilde{F_i} \) is normalized by its own mean value, \( \langle \tilde{F_i} \rangle \), yielding the normalized flat image \( \hat{F_i} = \frac{\tilde{F_i}}{\langle \tilde{F_i} \rangle} \). The mean flat image is computed as a weighted average of the normalized flat images, where the weights are determined by \( \langle \tilde{F_i} \rangle \), accounting for Poisson noise. Specifically, the mean flat image is given by

\[
\text{mean}(\hat{F}) = \frac{\sum_i \hat{F_i} \langle \tilde{F_i} \rangle}{\sum_i \langle \tilde{F_i} \rangle}.
\]

Additionally, the median flat image is computed as \( \text{median}(\hat{F}) \). The final master flat frame is then obtained by combining the median and mean flat images according to the following expression:

\[
F = 3 \, \text{median}(\hat{F}) - 2 \, \text{mean}(\hat{F}).
\]
This formulation of the stacked master flat frame helps mitigate the contamination from background stars in twilight flat images. It is an estimate of the mode of the flat field images, following Pearson's empirical formula for mode \citep{pearson1902}.

Calibrated science image $\hat{S_i}$ is 

\begin{equation}
    \hat{S_i} = \frac{S_i-B}{F}-\text{Sky}(\frac{S_i-B}{F}),
\end{equation}

where $S_i$ is the raw science image; $\text{Sky}()$ is the sky background obtained using \textsc{sep}\citep{Barbary2016}. Fluctuation of the background is also recorded in the database.

 \subsection{Stacked deep science image generation}
 \label{sec:stacking}
\begin{figure}
    \centering
    \includegraphics[width = 0.8\linewidth]{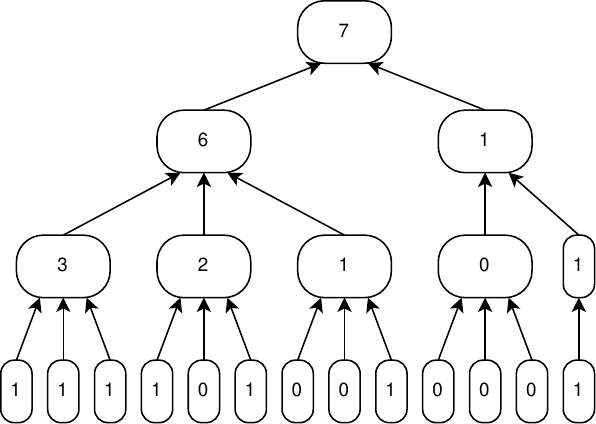}
    \caption{Demonstration of hierarchical stacking while stacking 13 images with fan-out equals to 3. A rectangle represents an image; number on the rectangles represents the total number of high-quality progenetor image; arrows point from the progenetor image to the stacked image. See Section \ref{sec:stacking} for details.}
    \label{fig:demostacking}
\end{figure}

 Stacking is a critical process in this pipeline, employed for multiple purposes. As aforementioned, stacking is used to generate the master bias and master flat images. Additionally, stacking is essential for extracting faint sources from deep science image when performing forced photometry. In the latter case, the stacking process becomes more complex. First, alignment is necessary due to the small deviations in the world coordinate system (WCS) of science images caused by tracking and pointing errors. Proper alignment is required to ensure that the images are stacked correctly. Furthermore, the stacking of masked images must be carefully considered. Aligning different images often results in blank regions, and the presence of masks for hot pixels can also affect the stacking process. Additionally, images with low quality, such as those affected by tracking issues, satellite or airplane tracks, or cloud cover, should be excluded from the stacking process. To address these considerations, we apply the following steps for deep science image stacking:

\begin{enumerate}
    \item \textbf{Resolving star and aligning images}: For each calibrated science image, source extraction is performed using \textsc{sep}. Translation between images is considered in the current pipeline. The deviation between two image $\Delta$ in direction $k\in \{x,y\}$ is 
    \begin{equation}
        \Delta_{k12} = \text{argmax}_n(\sum_{i,j} \mathds{1}\left[(k_{1i}-k_{2j})=n\right]),
    \end{equation}
    where $\Delta_{k12}$ is the deviation between image 1 and image 2 in direction $k$; $k_{1i}$ is the rounded position of the $i$th resolved star in the first image;  $\mathds{1}[\cdot]$ is the indicator function; $\text{argmax}_n(\cdot)$ is the $n$ such that the expression is at maximum. To ensure robustness, particularly in sparse star fields where the number of resolved star $i$ is small, and to avoid computationally expensive sub-pixel interpolation during stacking, the positions are rounded to integers, resulting in integer-valued $\Delta_k$.
    
    Meanwhile, number of resolved stars for each image is also recorded into the database.
    \item \textbf{Selecting image with high quality}: Low-quality images can result from multiple mechanisms. Unstable tracking may cause elongated star spots or even double images of a star. Cloud cover can obstruct part of the light, making stars harder to resolve. Airplane, satellite, and meteor tracks can introduce contamination, while images taken near twilight or morning may exhibit high background fluctuations. To filter out such contaminated images, two criteria are applied. First, a 3-sigma clipping is used on the number of resolved stars, with both upper and lower boundaries, to exclude images affected by cloud cover or tracking issues. Second, a 5-sigma upper boundary is applied to the sky background fluctuation, with no lower boundary, to reject images affected by tracks or high sun altitude angles. Both of these metrics are recorded in the aforementioned steps.
    \item \textbf{Stacking with given deviation} Given the relative translation of stacked image to the first image and the quality flag of an image, we would be able to stack the science image into a deep field image. A hierarchical tree structure in used in stacking pipeline. An demonstration of this algorithm is shown in Fig. \ref{fig:demostacking}. Each stacking is the weighted average of the progenetor images weighted by the number of high-quality images already stacked in the progenetor images. Process of stacking would begin automatically while all the processes of progenetor images are finished. These zero-weighted image with low quality flags are also considered in stacking because the process dependence relationship of the stacking can be generated before the image quality assessment process.

    By consuming more disk IO time, this stacking method can effectively save the RAM required for stacking. Meanwhile, it would enable extracting light curve with different time scales. Because median needs a global sorting, this algorithm does not work for master flat image generation. 
\end{enumerate}

To mitigate the rotation between different science images, high-accuracy WCS solution and resampling of the images are needed, which will be applied in future version of the pipeline. This would be essential for dealing with the image rotation problem in alt-azimuth mount of Tianyu. Meanwhile, it would increase the cross match accuracy of sources (defined in Appendix \ref{sec:concepts}).

\subsection{Cross match between known sources and new sources}

In this pipeline, source extraction is performed using \textsc{sep}. However, \textsc{sep} would not return the identity of sources. To enable transient detection, cross match between newly extracted sources and previously known sources is required. There are two types of previously known sources. One is the sources extracted from the same sky (defined in Appendix \ref{sec:concepts}), and another type is the sources from external catalogs. After sources are extracted from the template image, the pipeline would use a K-D tree to find the closest archive source in the database to the newly resolved sources. If the distance is larger than 1.5\,arcsec, it is registered as a new source. Meanwhile, in the current version of pipeline, Gaia is the only considered external catalog. \textbf{astrometry} \citep{Lang10} is used to obtain the WCS of the template image. Different input catalogs are used to ensure the success of astrometric resolution. Gaia sources in the vicinity of the target (defined in Appendix \ref{sec:concepts})is obtained using \textbf{astroquery} \citep{Ginsburg2019}. The projected position of Gaia sources are obtained using this WCS. In this pipeline, we record 3 closest Gaia sources and their distances to our sources. Meanwhile, the variability of the Gaia sources are also recoreded. It would be helpful for further source classification.

\subsection{Flux extraction and calibration}\label{sec:relativeflux}

In this pipeline, the flux extraction is performed with aperture photometry using \textsc{sep}. Apertures are placed on the position of resolved sources. Reference stars are grouped by position in the image and their flux quantiles.  Relative flux calibration requires nearby reference stars with similar magnitudes to the target stars. Meanwhile, systematic photometric error can be calibrated better when more reference stars are used. To balance the number of reference stars and width of flux bins, number of flux quantile bin $n_f$ in each position bin is set to be
\begin{equation}
    n_f = min(1,\lfloor N_f/N_{\text{min ref}}\rfloor,\sqrt{N_p}),
\end{equation}
where $N_p$ is the number of reference star candidate in this position bin; $N_{\text{min ref}}$ is the minimum number of reference star in this bin, which is set to be 6 in this pipeline; $\lfloor N_f/N_{\text{min ref}}\rfloor$ is the floor of $N_f/N_{\text{min ref}}$, which would guarantee the number of reference stars in each group.

Relative flux calibration within each reference group is performed using an algorithm similar to AstroimageJ \citep{Collins17}. For the $i$th non-reference star, the relative flux $\hat{F_i}$ can be obtained by
\begin{equation}
    \hat{F_i} = \frac{N_{\text{ref}}F_i}{\sum_{j\in\text{group}} F_{\text{r},j}},\label{eq:relfluxdef1}
\end{equation}
where $F_i$ is the raw flux of the $i$th star in the group; $F_{\text{r},j}$ is the $j$th reference star in this group; $N_{\text{ref}}$ is the number of reference star in this group. Meanwhile, for the $k$th non-reference star, the relative flux $\hat{F_k}$ is
\begin{equation}
    \hat{F_k} = \frac{(N_{\text{ref}}-1)F_k}{\sum_{j\in\text{group},j\ne k} F_{\text{r},j}}.\label{eq:relfluxdef2}
\end{equation}
Signal-to-noise ratio of the light curve in a group can be calculated by 
\begin{equation}
    {\rm SNR} = \frac{\sum_i \bar{\hat{F}}_j/\sigma_j^2 }{\sqrt{\sum_j \frac{1}{\sigma_j^2}}};\quad \sigma_j^2 = \frac{1}{n_t}\sum_t F_j(t)^2 - \bar{\hat{F}}_j^2;\quad \bar{\hat{F}}_j = \frac{1}{n_t}\sum_t \hat{F}_j(t)\label{eq:SNR}
\end{equation}
where $j$ enumerates all the stars in this reference group that have maximum number of epoch; $t$ is the epoch of observation; $n_t$ is the number of epoch. This signal-to-noise ratio would be used to select the reference stars.

 Reference stars are selected in the following way. A K-D tree is employed to identify the second closest resolved sources to each star. The resulting set consists of the closest sources, excluding the star itself. From this set, stars are selected based on their distance being within the upper 50\% quantile. Additionally, reference stars are present in every frame of the observation. Variable stars from the Gaia variable star catalog \citep{eyer2023} are also excluded from the selection. To avoid the situation where a resolved source is mistakenly identified as a single object composed of two nearby stars, the separation between the second nearest Gaia source and the closest matched Gaia source must be at least 1.5 times greater than the separation between the closest matched sources. These criteria would give a list of reference star candidates.

To further exclude the influence of unknown variability, e.g. stellar flare, of the reference stars, we applied an additional algorithm to select the reference stars from these candidates. Within each group, we calculate the signal-to-noise ratio of the light curves using the reference star candidates in this group using Eq. \ref{eq:SNR}. Meanwhile, we exclude the $k$th ($k\in [1,N_{\text{ref}}]$) reference star from the set of reference star candidates in this group, and calculate the signal-to-noise ratio again without this reference star. The $k$th reference star candidate in the group should be kept if signal-to-noise ratio is decreased after this star is being excluded from the reference star candidates.

\section{Relative photometry as a case study}
\label{sec:test}
\subsection{Description of data}
\label{sec:data}

Test data from Muguang Observatory (coordinate: 31$^\circ$10'07''N, 121$^\circ$ 36'21''E; altitude: 30 m) and Xinglong Observatory (coordinate: 40$^\circ$23'39''N, 117$^\circ$ 34'40''E; altitude: 960 m) are collected to test the performance of the software.

 Muguang Observatory is located at the rooftop of Tsung-Dao Lee Institute in Shanghai, China. A QHY600m CMOS camera \citep{alarcon23} is mounted on a CDK350 telescope system \footnote{\url{https://planewave.com/products/cdk350/}}. The equatorial tracking mode of CDK350 is used in this observation. The Baader CMOS L-Filter (420 – 685 nm) is used in this observation\footnote{\url{https://www.baader-planetarium.com/en/baader-uv-ir-cut-l-filter-cmos-optimized.html}}.

Meanwhile, we obtained the observation time for two nights of 85cm telescope in Xinglong Observatory on November 18-19th, 2024. Unfortunately, there is no observation in November 19th due to the weather condition. An Andor-DZ936N CCD is mounted on the telescope, which is mounted in an equatorial mount. The Johnson B-band filter is used in these observations. Details of the telescope are available in \url{http://www.xinglong-naoc.cn/html/en/gcyq/85/detail-26.html}.

Key parameters of the instruments used in this work is shown in Table \ref{tab:parins}. Because the observational systems of these two telescopes are not coupled into the architecture introduced in this paper, a program is used to move the image files into the file system given pre-registered observations. 

\begin{table}[!h]
    \centering
    \caption{Key parameters of the instruments.}
    \label{tab:parins}
    \begin{tabular}{lllllll}
    \hline
         Name&Diameter (cm)& Focal length (mm)&Camera type & Pixel scale ($\mu$m)& Pixel number&FoV (arcmin)\\\hline
         Muguang 35cm& 35&2563&QHY 600M &3.76&9576$\times$6388&48$\times$32\\
          Xinglong 85cm& 85&2987&Andor DZ936N BV &13.5&2048$\times$2048&32$\times$32\\\hline
    \end{tabular}
\end{table}

Two transit events are observed to test the performance of the architecture and relative photometry pipeline. Details of the observed targets are shown in Table \ref{tab:targets}. Configuration of the observation is available in Table \ref{tab:obsevents}.

\begin{table}[]
    \centering
    \caption{Parameters of observed targets. \citep{gaia23,guerrero21}}
    \label{tab:targets}
    \begin{tabular}{llllll}
    \hline
       Target name  &R. A. (deg) & Decl. (deg)&G-band magnitude & Transit depth (\%) & Transit duration (h)  \\\hline
        %HAT-P-20b &111.91642&24.33612&10.990$\pm$0.001&0.990&1.85$\pm$0.02\\
        %HAT-P-7b &292.24731&47.96950&10.368$\pm$0.003&0.696&	4.01$\pm$0.06\\
        TrES-5b &305.22196&59.44890&13.4599$\pm$0.0003&2.192&	1.60$\pm$0.05\\
        %Kepler-44b &300.10233&45.76219&14.6901$\pm$0.0002&0.740&3.12$\pm$0.01\\
        %Kepler-723b &284.83044&44.65809&15.2017$\pm$0.0005&1.785&3.230$\pm$0.002\\
        XO-4b &110.38804&58.26811&10.5131$\pm$0.0003&	0.880&4.45$\pm$0.06\\\hline
    \end{tabular}
\end{table}

\begin{table}[!h]
    \centering
    \caption{Observation profile}
    \label{tab:obsevents}
    \begin{tabular}{llllll}
    \hline
         Target name&Date& Instrument&Exposure time (s)&Binning & Number of image\\\hline
         %HAT-P-20b& Feb 16-17, 2024&Muguang 35cm&15&3&600\\
          %HAT-P-20b& Dec 8-9, 2024&Muguang 35cm&30&1&391\\
        %HAT-P-7b& Aug 7-8, 2024&Muguang 35cm&30&1&630\\
          TrES-5b& July 4-5, 2024&Muguang 35cm&30&1&522\\
          %Kepler-44b& Nov 18-19, 2024&Xinglong 85cm&30&1&218\\
          %Kepler-723b& Nov 18-19, 2024&Xinglong 85cm&30&1&100\\
          XO-4b& Nov 18-19, 2024&Xinglong 85cm&15&1&999\\
\hline
    \end{tabular}
\end{table}

\subsection{Experimental environment}
We applied the pipeline to the dataset described in Section \ref{sec:data}. The testing was conducted on a computer equipped with an Intel Core i7-11700K processor and a 1TB SSD. A UGREEN disk array configured in RAID0, comprising 5$\times$12TB hard drives, was connected to the test computer via a USB3.0 port. The file system was mounted on the disk array, while the database file resided on the computer's SSD. \textsc{MySQL} and \textsc{RabbitMQ} servers were deployed on this system, with the \textsc{RabbitMQ} server running within a Docker container. Since the observation control system is not yet developed, images and observation metadata were registered manually, with their \textit{process\_id}s serving as inputs to the pipeline.

\subsection{Performance}
\label{sec:scalability}

To evaluate the performance of the software, we record the resource consumption of different processes during the execution of consumers. The time and resources consumed by each step of the pipeline during the processing of XO-4 data using a single consumer are presented in Table \ref{tab:single_node}. The process that are called more than once, i.e. image calibration, alignment, stacking and flux extraction, can be executed in parallel by deploying more consumers, which would decrease the latency of light curve production of individual \textit{observation}. The other processes can be executed in parallel for different \textit{observation}s, which would increase the total throughput of the light curve production. The total execution time of the pipeline is 1490.28 s, including the overhead of process management. Meanwhile, 886,456 rows of  relative flux are extracted from 999 images. CPU and disk IO are the main bottlenecks of the performance, which can be solved by deploying more consumers and building a distributed storage system.

\begin{figure}
    \centering
    \includegraphics[height=0.8\linewidth]{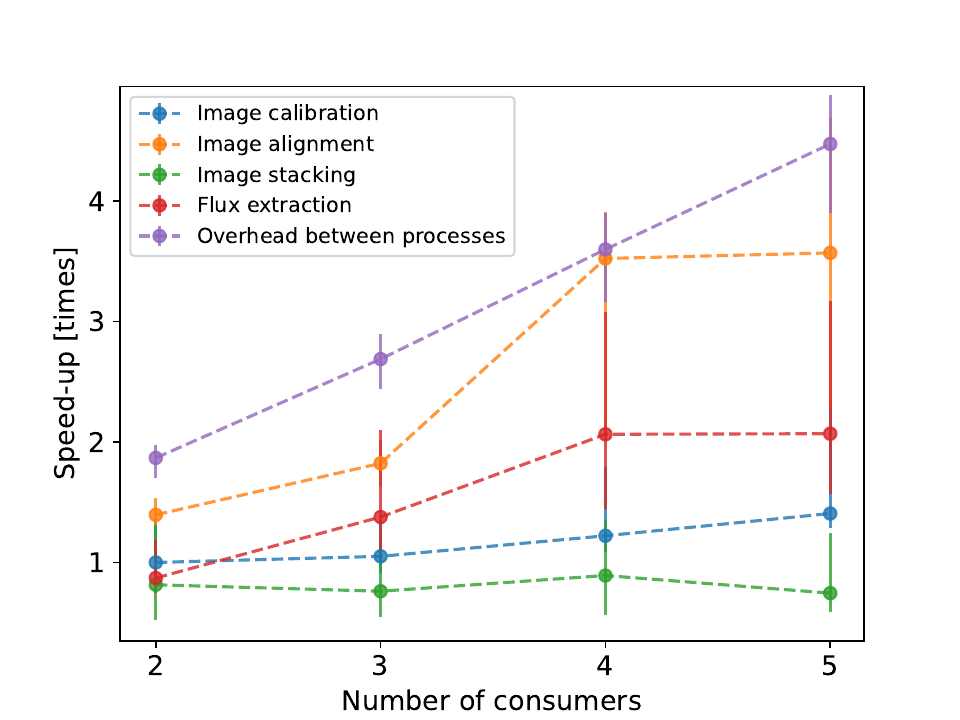}
    \caption{Relationship between the throughput speed-up relative to the single consumer and the number of deployed consumers. Consumers were deployed manually. Throughput values are obtained by the inverse execution time. The absolute total throughput as measured in s$^{-1}$ is available in Table \ref{tab:multi_node}.
    \label{fig:throughputres}}
\end{figure}

To evaluate the system's scalability, we measured the process throughput—defined as the number of processes completed by consumers per unit time—at various stages of the pipeline while varying the number of consumers. The total throughput of different pipeline components is shown in Table \ref{tab:multi_node}. The speed-up relative to using a single consumer is visualized in Fig. \ref{fig:throughputres}. As shown in the figure, throughput generally increases with the number of consumers, demonstrating the high scalability of the proposed architecture. Compared to a single consumer, the median throughput of image calibration, alignment, and flux extraction  increases by 41\%, 257\%, and 107\%  respectively when using 5 consumers.
However, we note that the total throughput of image calibration, stacking, and flux extraction does not increase significantly when the number of consumers increases from one to two. 
This is due to non-sequential read-out operations from the hard drive, which are less efficient. Additionally, the throughput of the stacking step does not scale proportionally with the number of consumers, as each consumer must read five images (the hierarchical stacking fan-out) and write one image. Meanwhile, the throughput of flux extraction and image alignment also converge when there are more than four consumers. These disk-intensive operations are ultimately constrained by disk I/O performance.

\begin{table}[!h]
    \caption{Time and resources consumed by each step of the pipeline using one consumer.  The data of XO-4 is used to assess the performance. The CPU, memory, and disk utilization values in the table correspond to the 16th, 50th, and 84th percentiles of resource usage during the execution of individual processes, with measurements taken once per second. Resources consumed by the MySQL server and the consumers are summed up. The CPU usage referes to a single thread of an intel chip. The execution time refers to the time consumed by a single call.}
    \centering
   %\begin{tabular}{llllllll}
\begin{tabular}{p{4.5cm}p{1cm}lllll}
\toprule
Process & Number of call & CPU (\%) & Memory (\%) & Read disks (kB/s) & Write disks (kB/s) & Execution time (s) \\
\midrule
Image calibration & 1004 & $97.00^{+7.00}_{-16.00}$ & $2.54^{+0.15}_{-0.03}$ & $24588.00^{+6971.68}_{-3076.00}$ & $49548.00^{+16412.00}_{-16396.48}$ & $0.28^{+0.09}_{-0.03}$ \\
Image alignment & 999 & $101.00^{+3.00}_{-4.00}$ & $2.67^{+0.01}_{-0.01}$ & $32777.00^{+1028.14}_{-110.72}$ & $100.00^{+64.00}_{-8.00}$ & $0.45^{+0.01}_{-0.01}$ \\
Image selection& 1 & $119.00^{+2.00}_{-62.96}$ & $2.68^{+0.00}_{-0.01}$ & $0.00^{+17.76}_{-0.00}$ & $0.00^{+4690.72}_{-0.00}$ & $10.71^{+0.00}_{-0.00}$ \\
Image stacking & 255 & $90.00^{+22.40}_{-15.00}$ & $3.21^{+0.36}_{-0.53}$ & $5896.75^{+74293.75}_{-5896.75}$ & $16866.00^{+16353.20}_{-262.00}$ & $0.56^{+0.37}_{-0.09}$ \\
Source detection& 1 & $96.00^{+5.52}_{-1.00}$ & $2.67^{+0.00}_{-0.00}$ & $0.00^{+32.00}_{-0.00}$ & $1520.00^{+27.04}_{-21.12}$ & $17.54$ \\
Cross match with Gaia & 1 & $102.00^{+2.00}_{-68.80}$ & $5.11^{+0.01}_{-0.82}$ & $468.00^{+34564.00}_{-468.00}$ & $612.00^{+5753.60}_{-612.00}$ & $26.56$ \\
Flux extraction & 999 & $92.00^{+8.00}_{-2.00}$ & $5.14^{+0.05}_{-0.00}$ & $298.50^{+465.82}_{-214.14}$ & $24204.00^{+5384.32}_{-5070.72}$ & $0.17^{+0.01}_{-0.02}$ \\
Relative flux calibration & 1 & $92.00^{+9.00}_{-29.92}$ & $6.01^{+0.00}_{-0.58}$ & $182.00^{+1454.80}_{-174.00}$ & $42878.00^{+13363.76}_{-33263.76}$ &  $262.48$ \\
Overhead between processes&3260&---&---&---&---&$0.021^{+0.002}_{-0.001}$\\
\bottomrule
\end{tabular}
    \label{tab:single_node}
\end{table}

\begin{table}[!h]
\caption{Total throughput of different processes with varying numbers of consumers. The values represent the 16th, 50th (median), and 84th percentiles of throughput for each task. All the values in the table are throughput in unit of per second.\label{tab:multi_node}}
\centering
\begin{tabular}{llllll}
\hline
 Number of consumers & 1 & 2 & 3 & 4 & 5 \\
\hline
 Image calibration & $3.51^{+3.87}_{-2.66}$ & $3.51^{+4.8}_{-2.8}$ & $3.69^{+4.74}_{-3.2}$ & $4.29^{+6.29}_{-3.82}$ & $4.94^{+8.18}_{-4.51}$ \\
  Image alignment & $2.20^{+2.27}_{-2.14}$ & $3.08^{+3.38}_{-2.85}$ & $4.02^{+4.44}_{-3.61}$ & $7.76^{+8.61}_{-4.61}$ & $7.86^{+10.33}_{-4.99}$ \\
  Image stacking & $1.79^{+2.12}_{-1.08}$ & $1.46^{+2.36}_{-0.94}$ & $1.37^{+2.21}_{-0.98}$ & $1.60^{+2.42}_{-1.02}$ & $1.34^{+2.23}_{-1.06}$ \\
  Flux extraction & $5.72^{+6.51}_{-5.48}$ & $4.99^{+6.78}_{-4.27}$ & $7.88^{+12.01}_{-6.11}$ & $11.8^{+17.61}_{-8.24}$ & $11.82^{+18.1}_{-8.96}$ \\
  Overhead between processes & $51.57^{+53.57}_{-46.99}$ & $96.34^{+101.87}_{-87.83}$ & $138.6^{+149.43}_{-125.96}$ & $185.49^{+201.26}_{-162.98}$ & $230.57^{+251.57}_{-201.09}$ \\
\hline
\end{tabular}
\end{table}

\subsection{Light curve and photometric precision}
\label{sec:lcapp}
\begin{figure}
    \centering
    \includegraphics[width = 0.45\linewidth]{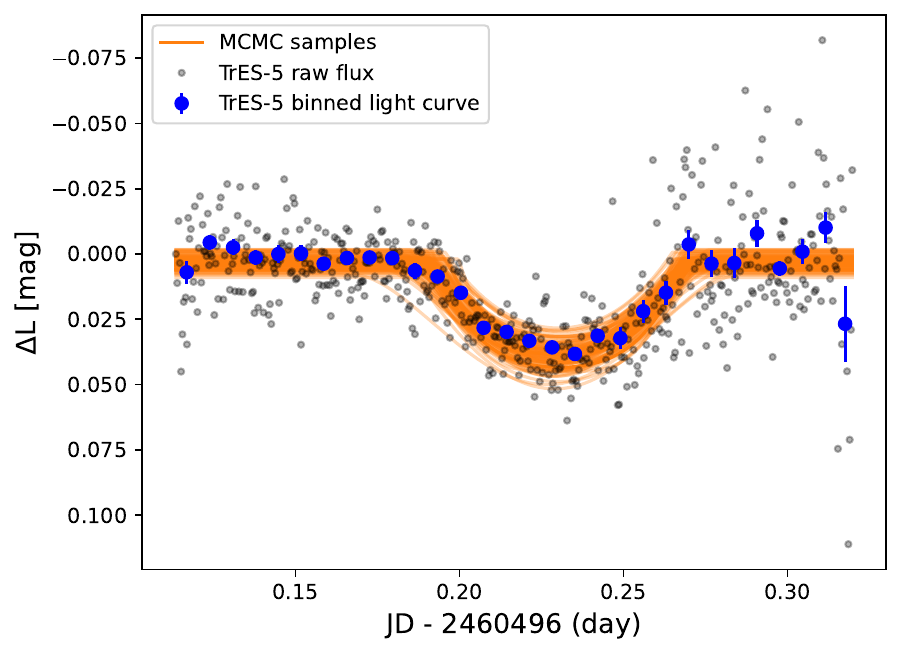}
     \includegraphics[width = 0.45\linewidth]{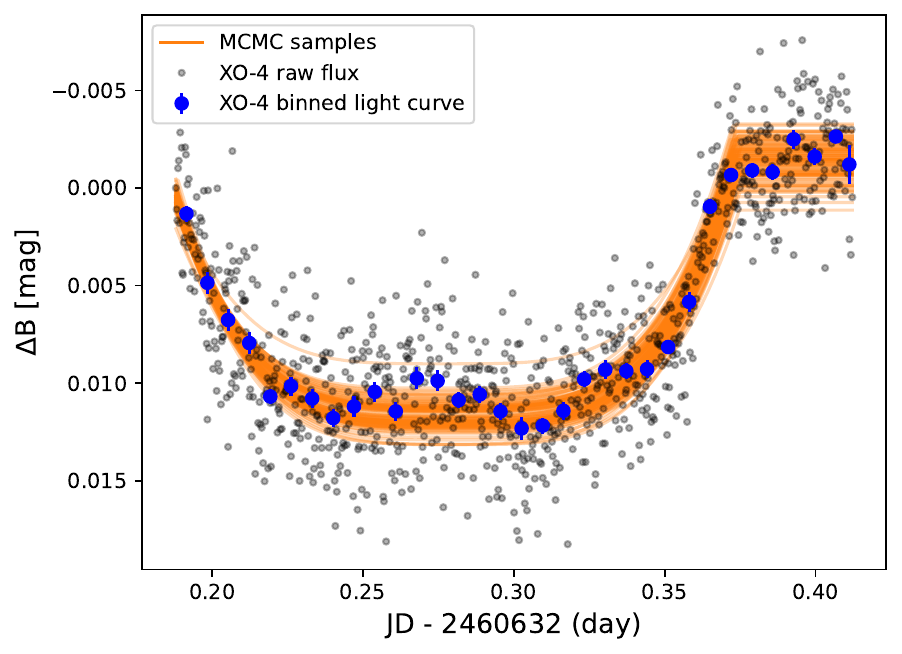}
    \caption{Relative flux of transiting targets. It is obtained using the proposed pipeline. Grey dots are the  relative flux extracted by the pipeline from the images that are used for template image generation;  blue points are the relative flux binned by 10 minutes; B is the filter used by Xinglong observatory and L is the filter used by Muguang observatory.  }
    \label{fig:lctransit}
\end{figure}

The detrended light curves of the transiting target produced by this pipeline are presented in Fig. \ref{fig:lctransit}. The relative flux is obtained using equation \ref{eq:relfluxdef1} and \ref{eq:relfluxdef2}. Details of the fitting, model comparison, and detending are  shown in Appendix \ref{sec:transitfitting}. The log Bayes Factor (lnBF) of the transiting signal of TrES-5 and XO-4 relative to the quadatic baseline model is 74.71 and 122.50 respectively, which indicate a strong transiting signal \citep{Kass1995}. %The relative flux of it decreases toward the end of the observation because of dawn light interference, which is successfully filtered out by the algorithm, as described in Section \ref{sec:stacking}. Additionally,  the end of XO-4's observation is affected by dawn sunlight, which is accurately recognized and addressed by the frame selection algorithm. 

\begin{figure}
    \centering
    \includegraphics[width = 0.3\linewidth]{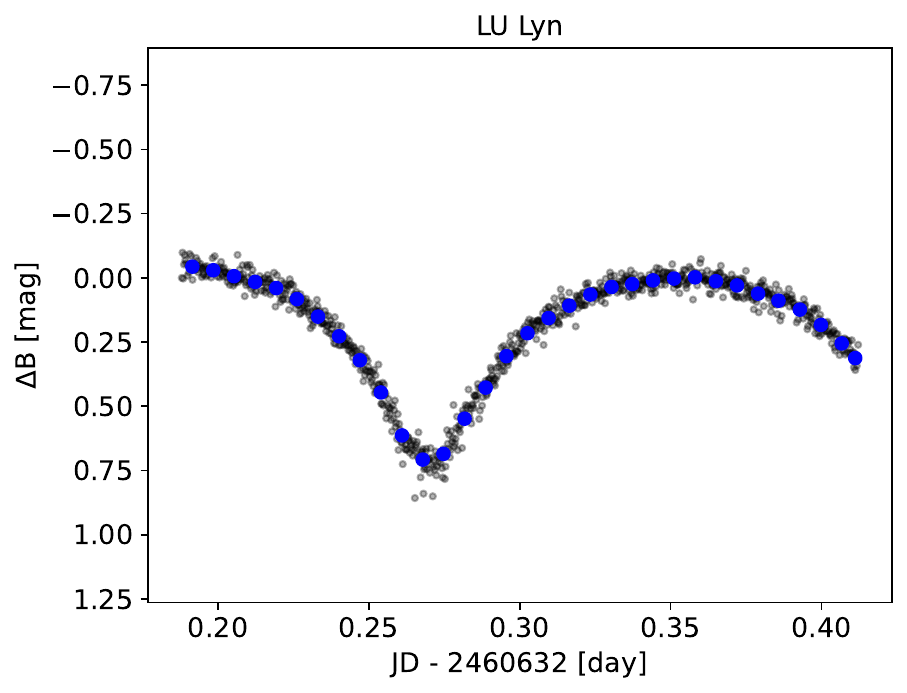}
    \includegraphics[width = 0.3\linewidth]{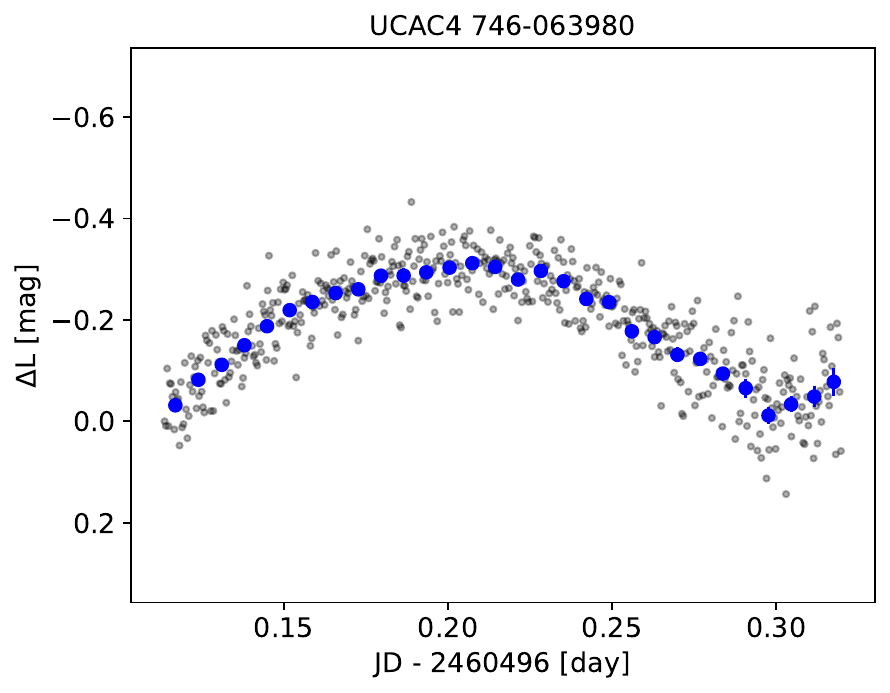}
    \includegraphics[width = 0.3\linewidth]{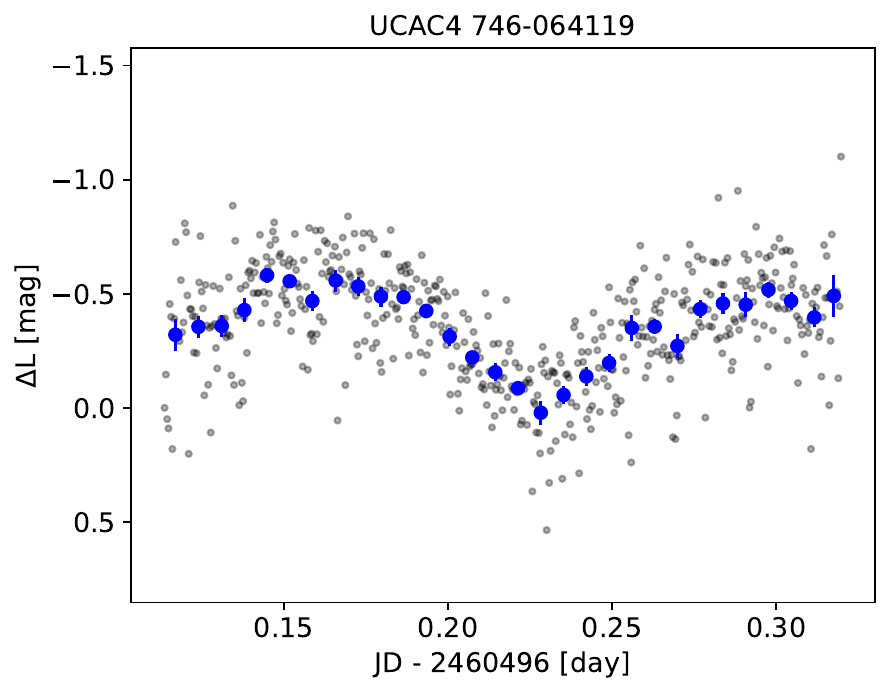}
    \includegraphics[width = 0.3\linewidth]{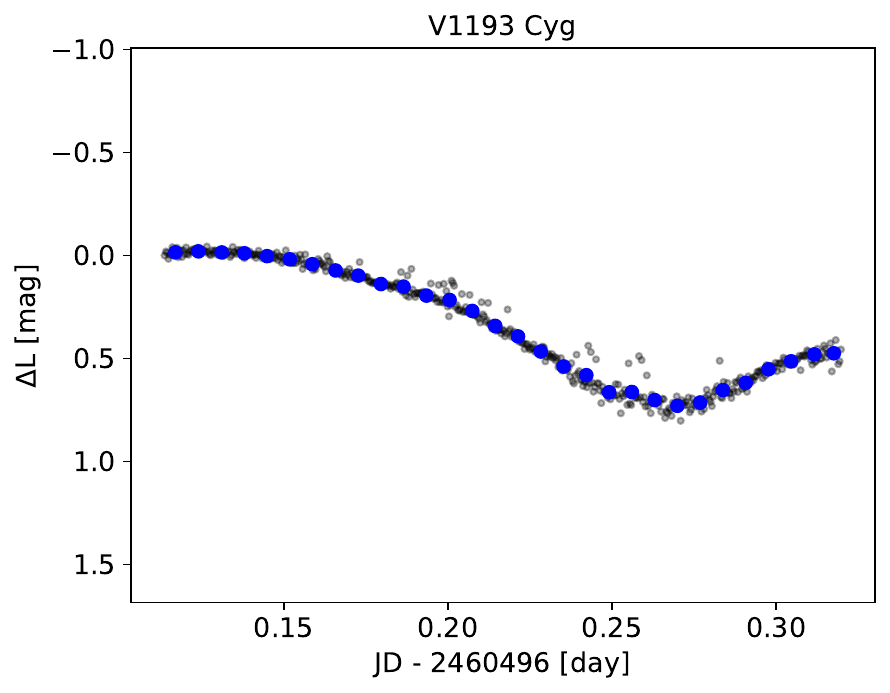}
    \includegraphics[width = 0.3\linewidth]{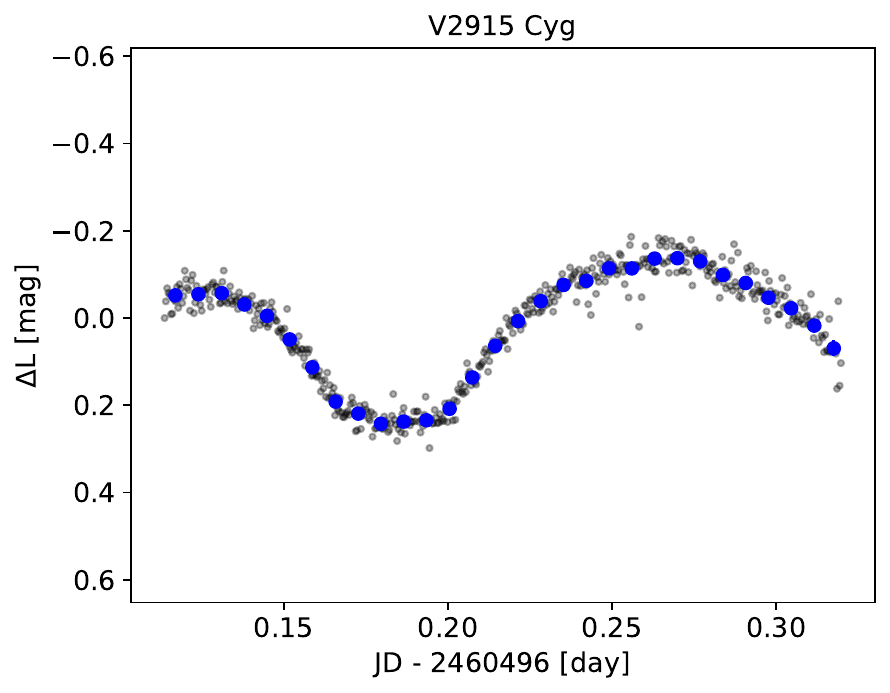}
    \includegraphics[width = 0.3\linewidth]{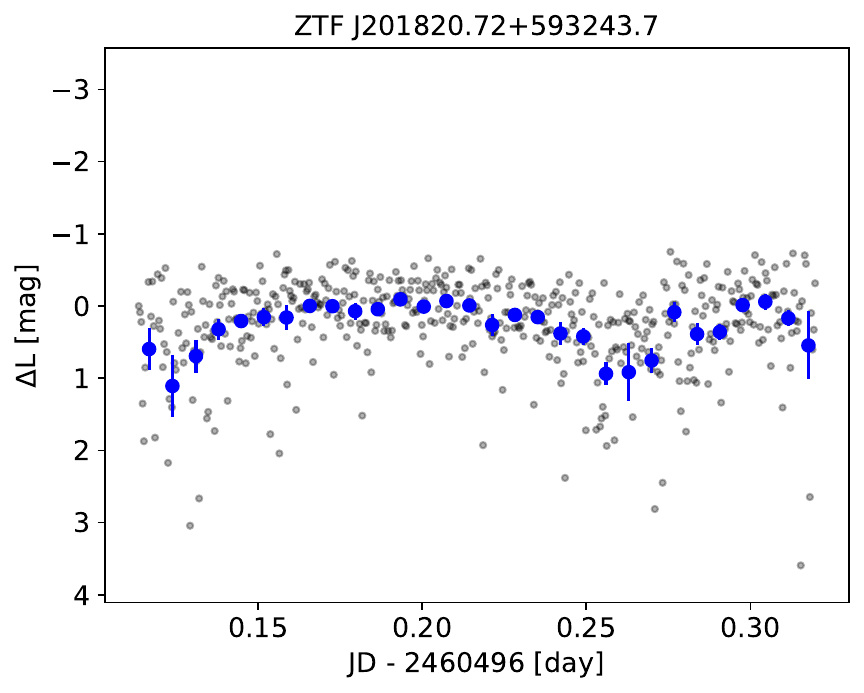}
    \includegraphics[width = 0.3\linewidth]{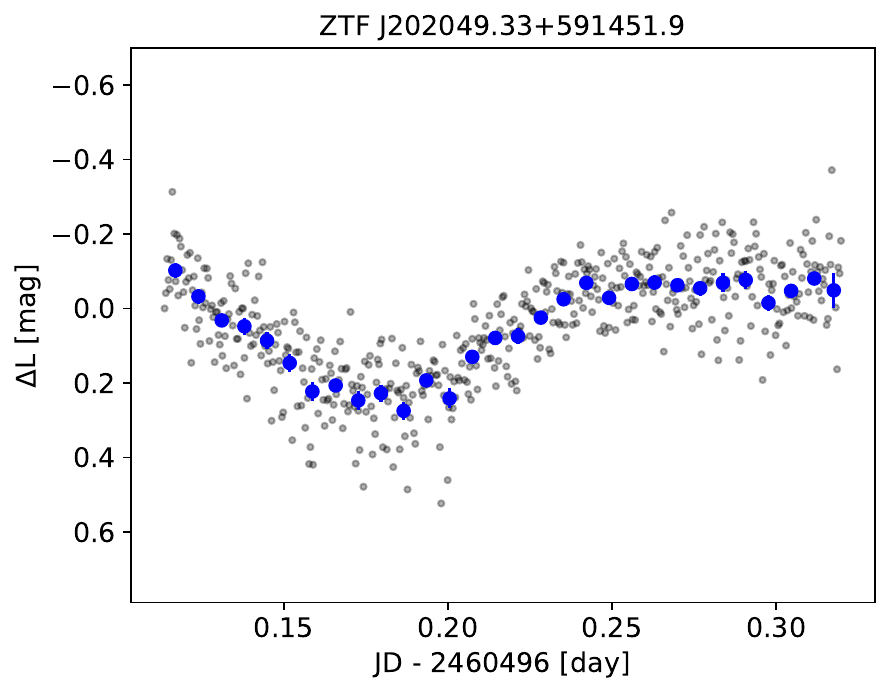}
    \includegraphics[width = 0.3\linewidth]{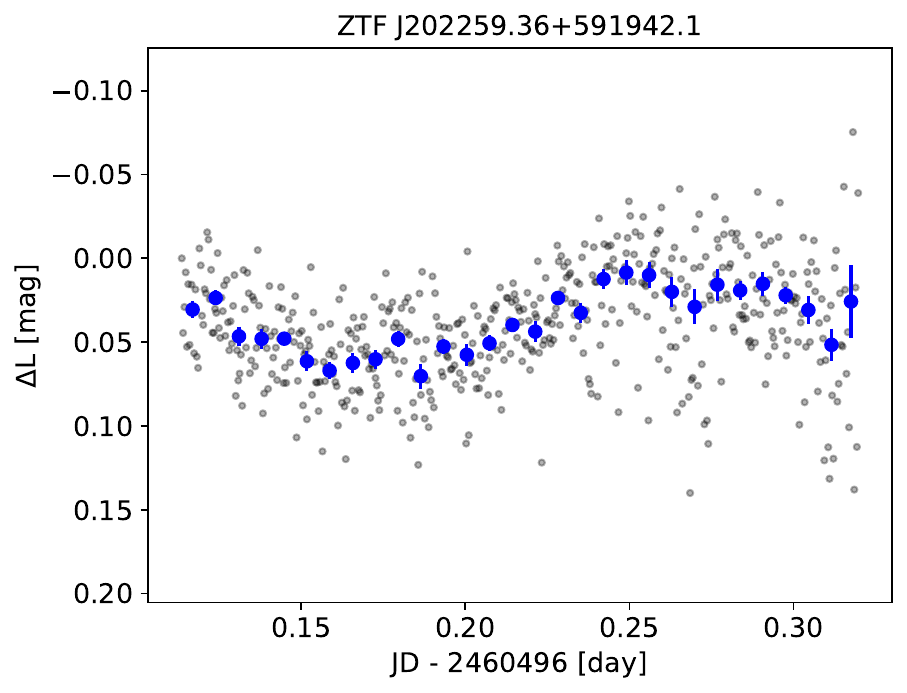}
    \includegraphics[width = 0.3\linewidth]{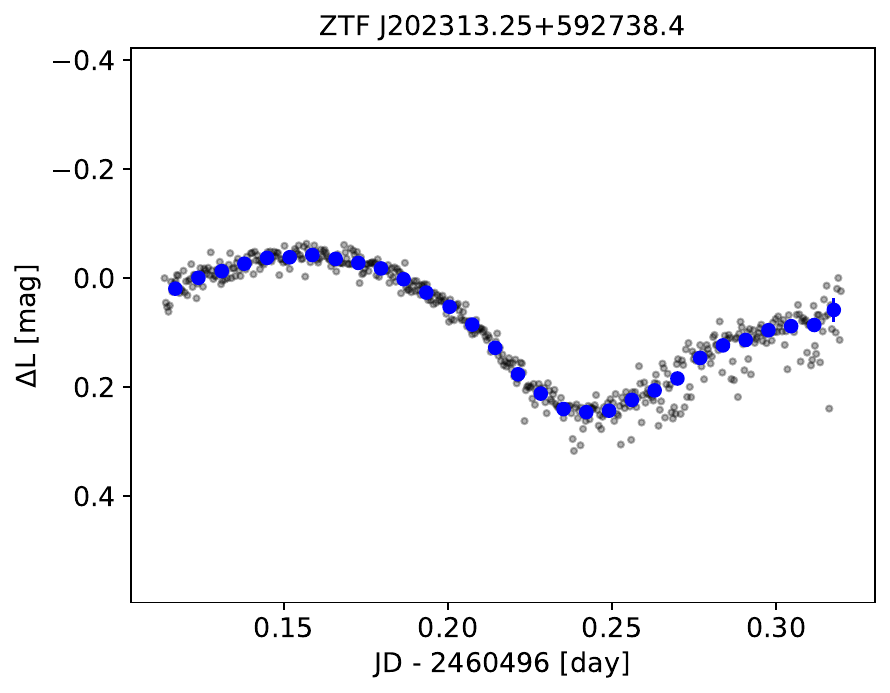}
    \caption{Relative flux of known variable stars. The configuration is identical to that in Fig. \ref{fig:lctransit}. The B-band data were obtained using the Xinglong 85 cm telescope, while the L-band data were acquired using the Muguang 35 cm telescope.}
    \label{fig:lcknownvariable}
\end{figure}

\begin{figure}
    \centering
    \includegraphics[width = 0.45\linewidth]{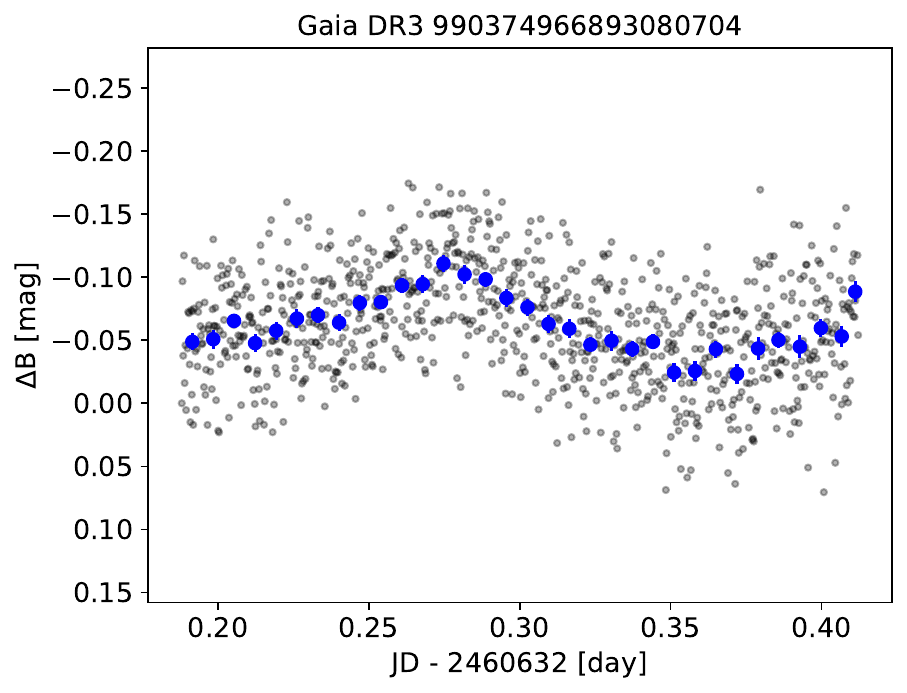}
    \includegraphics[width = 0.45\linewidth]{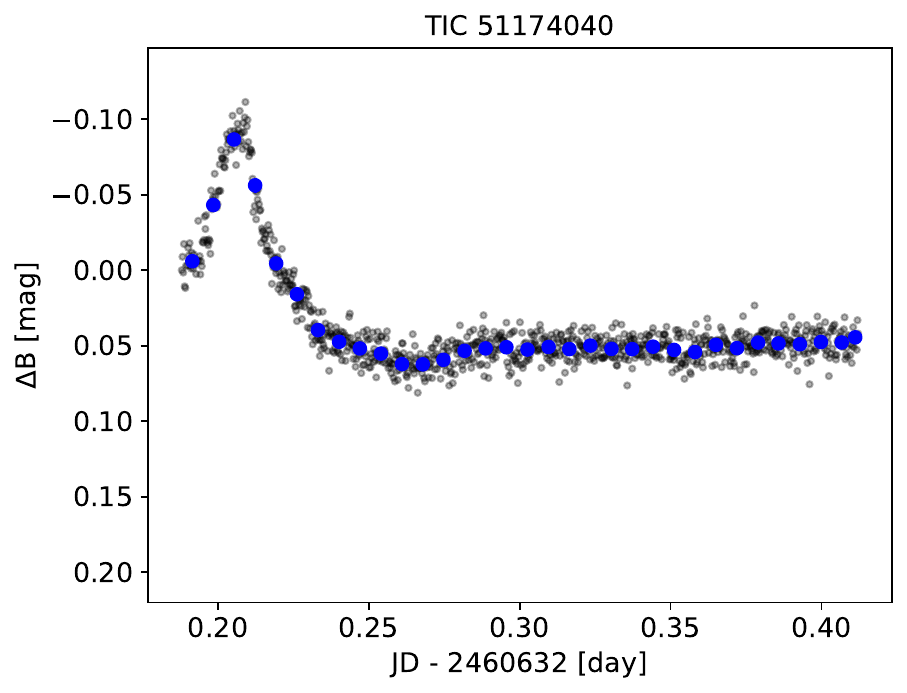}
    \caption{Relative flux of unreported variable sources. Configuration is the same as Fig. \ref{fig:lctransit}.}
    \label{fig:lcnewvariable}
\end{figure}

The cross-match with the Gaia variable star catalog \citep{eyer2023} confirms the presence of several known variable stars near the observed transiting targets. The relative fluxes of these known eclipsing binaries are displayed in Fig. \ref{fig:lcknownvariable}, with their flux variations closely matching the measurements reported in the Gaia DR3 variable star catalog. In addition to these known variables, several new variable sources have been identified, as shown in Fig. \ref{fig:lcnewvariable}. The light curve morphology and stellar parameters from Gaia DR3 suggest that Gaia DR3 990374966893080704 is likely an RR Lyrae variable star, while  TIC 51174040 is consistent with stellar flare events. We need to notice that TIC 51174040 is one of the reference star candidates in this observation, which is ruled out using the algorithm in Section \ref{sec:relativeflux}.

These light curves demonstrate the ability of the proposed architecture and pipeline to produce data suitable for detecting transiting giant planets and variables. Furthermore, they highlight the pipeline's ability for processing data from multiple instruments.

%red dots are the relative flux extracted from the images that are excluded from the template image stacking, as described in Section \ref{sec:stacking};

\begin{figure}
    \centering
    \includegraphics[width = \linewidth]{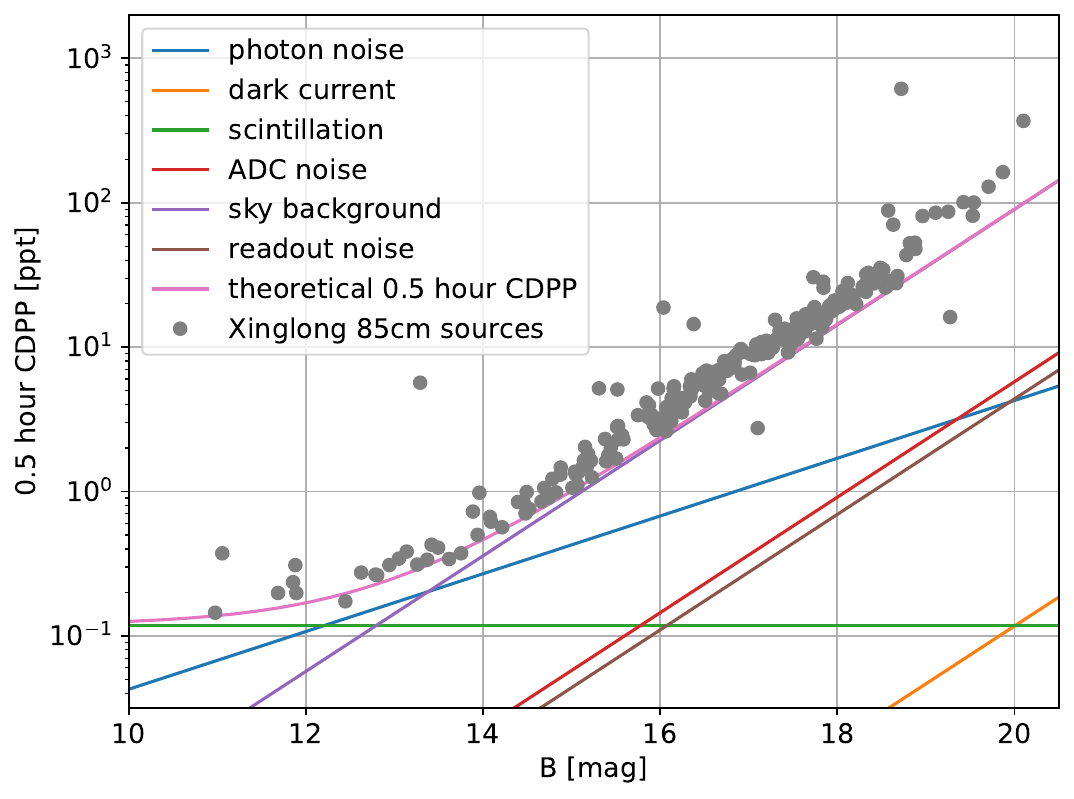}
    \caption{CDPP of the sources in the vicinity of XO-4 compared with the theoretical CDPP. Different straight lines represents the noise from different components. See Section \ref{sec:lcapp} for details.}
    \label{fig:cdpp}
\end{figure}

To qualitatively assess the photometric precision of the pipeline, we calculate the combined differential photometric precision (CDPP) for XO-4 observations using \textsc{lightkurve} \citep{lightkurve13}. An 1-hour timescale detrending based on Savitzky–Golay filter is applied to the relative flux, and the CDPP at a 0.5-hour timescale is computed from the detrended light curves of the target stars. The B-band magnitudes of the stars observed with the Xinglong 85 cm telescope are obtained using the  BEst STar database (Xiao et al., in prep.). It provides a 5-band standard star catalog in the Xinglong 85 telescope system, encompassing approximately 200 million ``corrected'' Gaia BP/RP (XP) spectra \citep{Huang2024} based synthetic photometric  (XPSP) standard stars with internal precision less than 0.01 mag derived using an improved XPSP method \citep{Xiao2023}.
 The method for calculating the theoretical CDPP, along with an interpretation of its different components, is described in \citealt{Feng24}. Instrumental parameters used for the CDPP calculation are available on the official website\footnote{\url{http://www.xinglong-naoc.cn/html/en/gcyq/85/detail-26.html}}. The sky background is assumed to be 18 $\rm mag/arcsec^2$, based on the moon phase on November 18, 2024. The results are shown in Fig. \ref{fig:cdpp}. The lower boundary of the measured CDPP for the extracted relative flux aligns closely with the theoretical prediction, demonstrating the pipeline's capability to produce high-accuracy light curves.

\section{Conclusion and discussion}
\label{sec:condis}

The architecture of Tianyu software is designed for high scalability, automation, and efficiency, enabling large-scale photometric data reduction—an essential feature for modern optical survey telescopes. Its fully automated process management efficiently handles the vast data generated by the Tianyu telescope while maintaining flexibility for future expansions.

Performance evaluation using observational data from the Muguang and Xinglong Observatories confirms the system's scalability. Compared to a single consumer, the median throughput of image calibration, alignment, and flux extraction increases by 41\%, 257\%, and 107\% respectively when using 5 consumers. However, image stacking exhibits limited scalability due to disk I/O constraints, with throughput decreasing when deploying multiple consumers.  The extraction of 886,456 rows of relative flux from 999 images in the XO-4 dataset was completed in 1490.28 s using a single consumer. Disk I/O and CPU usage were identified as the primary bottlenecks.

Relative photometry tests validate the accuracy of the pipeline, with differential photometric precision approaching theoretical limits. The combined differential photometric precision (CDPP) of XO-4 observations for a B$=17$ star is measured at 1\% for a 0.5-hour timescale, aligning closely with the theoratical limit. The system successfully detected two transiting exoplanets with log Bayes factors of 74.71 and 122.50, confirming strong transit signals. Additionally, the pipeline identified variability in nine known variable stars and two previously unreported sources, demonstrating its capability for high-precision time-domain astronomy.

Compared to general-purpose photometric software, the Tianyu pipeline introduces significant improvements in batch processing, automation, parallelism, memory efficiency, and result management. These advancements make it well-suited for large-scale surveys, reducing human intervention while improving computational performance. Furthermore, the robust database-driven approach ensures seamless data retrieval and integration with external catalogs, facilitating long-term monitoring of variable sources.

Future development of the Tianyu pipeline will focus on further reducing latency, enhancing debugging capabilities, and extending its functionality for telescope control, absolute flux calibration, variable source detection and differential imaging. In the future, we also plan to evaluate sub-pixel registration methods, such as the SExtractor \citep{bertin96} –SCAMP \citep{bertin06}–SWARP \citep{bertin10} toolchain, to assess their suitability for Tianyu's critically sampled images.  These enhancements will improve the pipeline’s applicability to a broader range of astrophysical studies, including transient event detection and photometric follow-up of exoplanet candidates.

By laying the groundwork for an efficient and extensible data processing system, this work ensures that the Tianyu telescope will serve as a valuable instrument for exoplanet discovery and time-domain astronomy, contributing to our understanding of dynamic celestial phenomena.

\appendix    %>>>> this command starts appendixes
 
\section{Concepts}
\label{sec:concepts}
We define several abstract concepts to illustrate how does Tianyu software work, which are shown in Table \ref{tab:concepts}.

\begin{table}[h!]
\caption{Descriptions of Key Concepts\label{tab:concepts}}
\centering
\begin{tabular}{|l|p{12cm}|} % Adjust the width as needed
%\begin{tabular}{|l|l|}
\hline
\textbf{Item} & \textbf{Description} \\ \hline
\textbf{Process} & A process refers to any behavior executed by a computer, including simulated image generation, image stacking, light curve extraction, data transfer, mount control, camera capture, and data compression. Each process has a unique \textit{process\_id} (PID) recorded in the database, consisting of a timestamp and a 6-digit random number. Dependencies between processes are also recorded. \\ \hline
\textbf{Site and Group} & The same site shares the same IP address and file system, while the same group shares hardware resources. For example, one group of consumers is linked directly to the telescope and camera, while another group has access to GPU resources. \\ \hline
\textbf{Target} & A target is the object of observation, such as a calibration image (e.g., flat field), a single star, a moving object like an asteroid, or a fixed sky region for survey. \\ \hline
\textbf{Observation} & An observation refers to a unit where images with the same target are processed together. For example, flat field images taken at twilight belong to one observation, and images of a fixed sky region captured with a 30-minute cadence belong to another observation. \\ \hline
\textbf{Sky} & A sky has fixed first-order World Coordinate System (WCS) information (e.g., right ascension, declination, field of view). An observation may not necessarily have a sky (e.g., for bias), and multiple observations can target the same sky. A target may correspond to multiple skies, depending on instrumentation and rotator configuration. \\ \hline
\textbf{Source} & A source is extracted from a fixed position in the template image of a sky and is recognized by its pixel position in the image. A star in overlapping sky regions can be recognized as two sources, and a moving object in a fixed sky region may be recognized as multiple sources. Cross-matching with external catalogs is required for identification. \\ \hline
\end{tabular}
\end{table}

\section{Database Schema}
\label{sec:dbs}

\begin{figure}
    \centering
    \includegraphics[width=\linewidth]{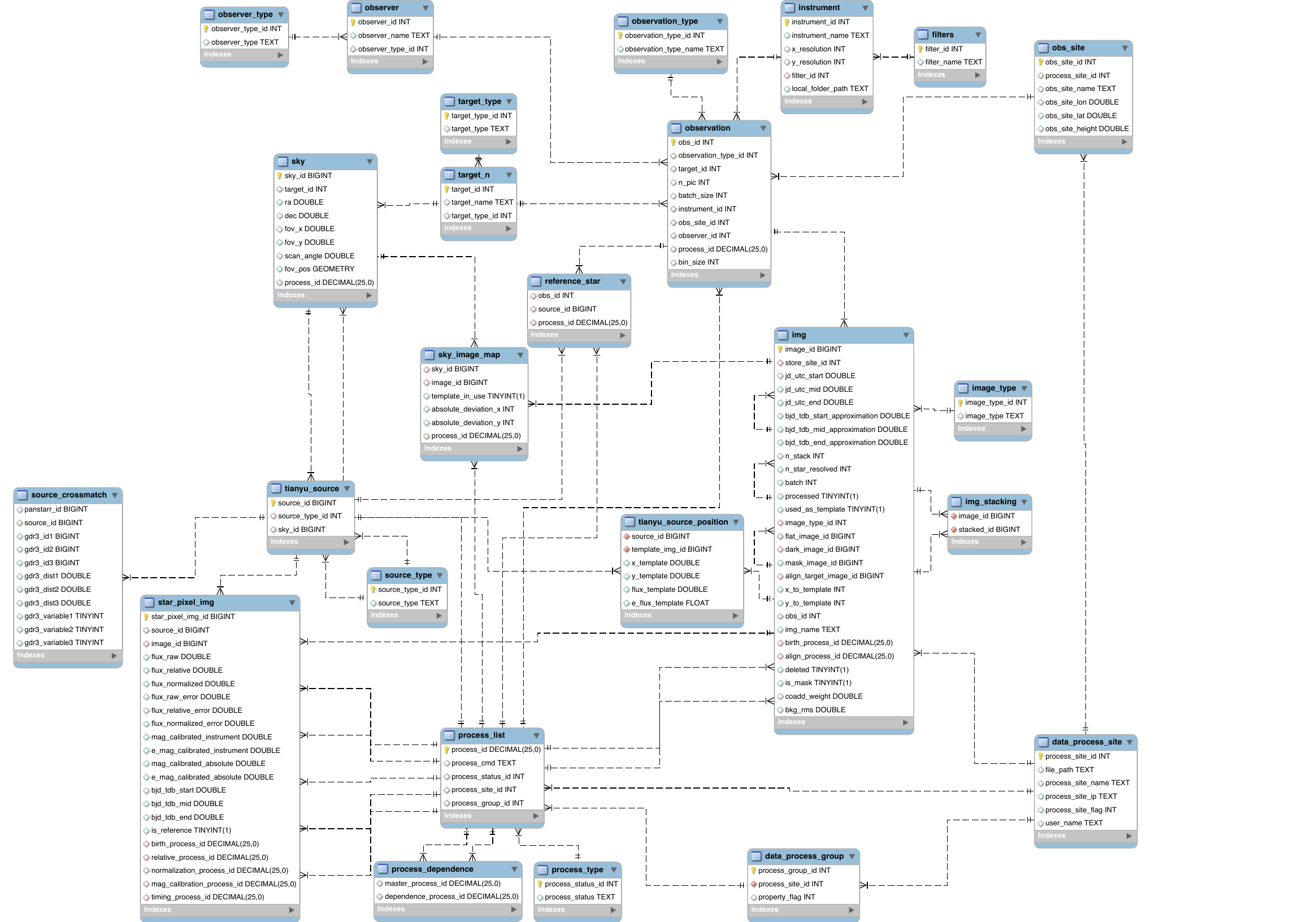}
    \caption{Entity-Relation diagram  of Tianyu database. A relational schema comprising multiple tables is depicted, with dashed lines connecting entities to represent relationships between them. The direction of these connections is denoted by the arrowhead, which serves as a foreign key referencing the corresponding table column in the related entity.  }
    \label{fig:dbdiag}
\end{figure}

As mentioned in Section \ref{sec:database}, the database schema is not totally fixed because the software is still uder development. Therefore, we only show the schema of key tables that would not have (significant) changes in the future.  The current database schema is shown in Fig. \ref{fig:dbdiag}. \textsc{Mysql Workbench}\footnote{\url{https://www.mysql.com/products/workbench/}}is used for this visualisation. Description of physical schema are available in \url{https://github.com/ruiyicheng/Tianyu_pipeline/blob/main/schema_design.md}. 

\section{Single-transit fitting and model comparison}
\label{sec:transitfitting}

In section \ref{sec:lcapp}, the following model is used to fit the single-transit light curve of TrES-5 and XO-4. The likelihood of the single-transit model is 
\begin{equation}
    \begin{aligned}
    &P(f|\text{transit},A_2,A_4,B_0,B_1,B_2,\sigma_f,t_0,d) \propto  {\rm Prior} (A_2,A_4)\cdot \mathcal{N}(f-X_{\rm t}\vec{\theta}_{\text{t}},\sigma_f^2 I),\quad \vec{\theta}_{\text{t}} = \begin{pmatrix}A_2,A_4,B_0,B_1,B_2\end{pmatrix}^T\\ 
    &X_{\rm t} = \begin{pmatrix}
        x_1^2&x_1^4&1&t_1^{'}&t_1^{'2}\\
        x_2^2&x_2^4&1&t_2^{'}&t_2^{'2}\\
        \vdots&\vdots&\vdots&\vdots&\vdots\\
        x_N^2&x_N^4&1&t_N^{'}&t_N^{'2}
    \end{pmatrix},\quad x_i = \left\{\begin{aligned}
       t_i^{'} & \quad -1<t_i<1 \\
       1 & \quad \text{elsewise}
    \end{aligned}\right .,\quad t_i^{'} = \frac{2(t_i-t_0)}{d},\\& {\rm Prior} (A_2,A_4) = \mathds{1}[A_2<0 \text{ and } (A_4<0 \text{ or } -\frac{A_2}{2A_4}>1)].
\end{aligned}
\end{equation}In this model, \( f \) represents the relative flux of the light curve, measured as a change in magnitude. The vector \( \vec{\theta}_{\text{t}} \) contains the linear parameters for the transit model, which include terms for the transit itself (\( A_2, A_4 \)) and for the intra-night trend (\( B_0, B_1, B_2 \)). \( \sigma_f \) denotes the white noise of the flux, while \( d \) is the transit duration, and \( N \) is the number of photometric points. \( I \) is an \( N \times N \) identity matrix, and \( t_i \) represents the epoch of the \( i \)-th observation. $\mathcal{N}(f-X_{\rm t}\vec{\theta}_{\text{t}},\sigma_f^2 I)$ represents the multivariate normal distribution with mean vector $f-X_{\rm t}\vec{\theta}_{\text{t}}$ and covariance matrix $\sigma_f^2 I$. The normalized time \( t_i^{'} \) is defined as the observational time relative to the mid-transit time \( t_0 \) and scaled by the transit duration \( d \). A prior is applied to ensure that the flux decreases monotonically during the ingress and increases monotonically during the egress of the transit. The intra-night trend, obtained by multiplying the last three columns of $X_t$ with the last three rows of $\vec{\theta}_t$, is subtracted in Fig. \ref{fig:lctransit}.
 Meanwhile, the baseline model with no transit is 
\begin{equation}
    P(f|\text{no transit},B_0,B_1,B_2,\sigma_f,t_0,d) =  \mathcal{N}(f-X_{\rm nt}\vec{\theta}_{\text{nt}},\sigma_{f,nt}^2 I),\quad \vec{\theta}_{\text{nt}} = \begin{pmatrix}B_0,B_1,B_2\end{pmatrix}^T,\quad
    X = \begin{pmatrix}
        1&t_1&t_1^2\\
        1&t_2&t_2^2\\
        \vdots&\vdots&\vdots\\
        1&t_N&t_N^{2}
    \end{pmatrix},
\end{equation}.

\textsc{Emcee} \citep{foreman13} is used as the Markov chain Monte Carlo (MCMC) sampler for the parameter estimation. Using the posterior samples, the lnBF for single-transit model relative to quadratic baseline model can be calculated by 
\begin{equation}
\begin{aligned}
    {\rm lnBF} &= \ln P(f|{\rm transit}) - \ln P(f|\text{no transit})\\
    &\approx \ln \left(\frac{1}{M}\sum_j^M P(f|\text{transit},\vec{\theta}_{{\rm t}j},\sigma_{fj},t_{0j},d_j)  \right) - \ln \left(\frac{1}{M}\sum_j^M P(f|\text{no transit},\vec{\theta}_{{\rm nt}j},\sigma_{f,{\rm nt}j})  \right)
\end{aligned}
\end{equation},
where $M$ is the number of posterior sample; $j$ represents the $j$th sample. The equation $\ln\left(\sum_{i=1}^N e^{l_i}\right) = \ln\left(\sum_{i=1}^N e^{l_i - \max(l)}\right) + \max(l)$ is used to avoid numerical overflow.

\section*{ACKNOWLEDGMENTS}  % equivalent to \section*{ACKNOWLEDGMENTS}       
 
We would like to thank Xianzhong Zheng for useful discussion. We would like to thank Haofeng Liu for his assistance in network technology.
 This work is supported by the National Key R\&D Program of China, No. 2024YFA1611801 and No. 2024YFC2207800, by the National Natural Science Foundation of China (NSFC) under Grant No. 12473066, by the Shanghai Jiao Tong University 2030 Initiative.
K.X. acknowledge the supports of the NSFC grant No. 12403024, the Postdoctoral Fellowship Program of CPSF under Grant Number GZB20240731, the Young Data Scientist Project of the National Astronomical Data Center, and the China Post-doctoral Science Foundation No. 2023M743447.
We acknowledge the support of the staff of the Xinglong 85cm telescope. This work was partially supported by the Open Project Program of the Key Laboratory of Optical Astronomy, National Astronomical Observatories, Chinese Academy of Sciences. This project was partially supported by Office of Science and Technology, Shanghai Municipal Government (grant Nos. 23JC1410200, ZJ2023-ZD-003).

\vspace{5mm}
\facilities{Xinglong 85cm, Muguang Observatory}
\software{MySQL (\url{https://www.mysql.com/}), RabbitMQ (\url{https://www.rabbitmq.com/}), sep \citep{Barbary2016}, numpy \citep{Harris20}, pandas \citep{reback2020pandas}, astropy \citep{astropy13}, astroquery \citep{Ginsburg2019}, astrometry \citep{Lang10}, lightkurve \citep{lightkurve13}, matplotlib \citep{Hunter2007}, emcee \citep{foreman13}, corner \citep{corner}}


\bibliography{PASPsample631}{}
\bibliographystyle{aasjournal}



\end{document}